\RequirePackage{lineno}
\documentclass[aps,prl,superscriptaddress,onecolumn,notitlepage]{revtex4-1}

\usepackage{graphicx}
\usepackage{CJK} 
\usepackage{amsmath}
 \usepackage{amssymb}
\usepackage{epstopdf}
\usepackage{color}
\usepackage{float}
\usepackage{bm}
\usepackage[normalem]{ulem}

\begin{document}
		\title{Ice front shaping by upward convective current}
		
		\author{Ziqi Wang}
		\affiliation{Center for Combustion Energy, Key laboratory for Thermal Science and Power Engineering of Ministry of Education, Department of Energy and Power Engineering, Tsinghua University, Beijing 100084, China}			
			
		\author{Linfeng Jiang} 
		\affiliation{Center for Combustion Energy, Key laboratory for Thermal Science and Power Engineering of Ministry of Education, Department of Energy and Power Engineering, Tsinghua University, Beijing 100084, China}			
		
		\author{Yihong Du} 
		\affiliation{Center for Combustion Energy, Key laboratory for Thermal Science and Power Engineering of Ministry of Education, Department of Energy and Power Engineering, Tsinghua University, Beijing 100084, China}	
			
		\author{Chao Sun}
		\thanks{chaosun@tsinghua.edu.cn}
		\affiliation{Center for Combustion Energy, Key laboratory for Thermal Science and Power Engineering of Ministry of Education, Department of Energy and Power Engineering, Tsinghua University, Beijing 100084, China}
		\affiliation{Department of Engineering Mechanics, School of Aerospace Engineering, Tsinghua University, Beijing 100084, China}
		
		\author{Enrico Calzavarini} 
			\thanks{enrico.calzavarini@univ-lille.fr}
		\affiliation{Univ.\ Lille, Unit\'e de M\'ecanique de Lille - J. Boussinesq - UML - ULR 7512, F-59000 Lille, France}
			
		\date{\today}

		\begin{abstract}
The extent and the morphology of ice forming in a differentially heated cavity filled with water are studied by means of experiments and numerical simulations. We show that the main mechanism responsible for the ice shaping is the existence of a cold upward convective current in the system. Such a current is ascribed to the peculiar equation of state of water, i.e., the non-monotonous dependence of density with temperature. The precise form of the ice front depends on several factors, first the temperature difference across the cell which drives the convection, second the wall inclination with respect to the vertical, both of which are here explored. We propose a boundary-layer model and a buoyancy-intensity model which account for the main features of the ice morphology.
		\end{abstract}
	\maketitle

Turbulent convective flows along with the ice formation process create intriguing coupling behaviors, which have a widespread appearance in nature and strong relevance in industrial applications \cite{meakin2010geological,Alboussiere2010Melting,epstein1983complex,worster1997convection,scagliarini2020modelling}. 
In general, the orientation of the temperature gradient and the gravity vector are not parallel. Their angle can play an important role in determining the ice front morphology and system heat transfer performances. Examples are the surficial icing of lakes and rivers, floating ice bodies (e.g., icebergs), ice bodies (e.g., ice shelf) extending outward from the land into waters, and solidification in energy storage technology \cite{huppert1978melting, russell1980melting, rignot2013ice, kousha2017effect, pogorelova2019moving,hester2021aspect}. For water, the coupled physics among the phase change, turbulent convection, and the density anomaly (water density reaches a maximum, $\rho_c$, at the density peak temperature, $T_\text{c}$ ($\approx 4^\circ$C)) bring more challenges: the gravitationally stable and unstable stratification coexist, which can strongly affect flow structures \cite{malm1998bottom,kowalewski1999freezing,Veronis1963Penetrative,arid2012numerical,large2014penetrative,corcione2015penetrative, ayurzana2016phase,2020_coupled,lecoanet2015numerical,toppaladoddi2018penetrative,dietsche1985influence,hu2017lattice}. 

In recent years, many studies are devoted to exploring the interplay between convective flows and thermal stratification/phase-transition under different system inclination. For penetrative convection \cite{Veronis1963Penetrative}, some of these have found that the inclination can induce the breakdown of the fluid stratification due to density anomaly \cite{inaba1984natural, quintino2017heat,quintino2018optimal}. Others have explored the coupling dynamics of phase-change and turbulent convection using phase-change material 
\cite{kamkari2014experimental,kamkari2017numerical,zeng2017numerical,madruga2021effect}.  
A recent work \cite{wang2020growth}, in a freezing-from-above system, showed that the density anomaly induced stratification has major effects on the flow structures and the resulting ice front speed and equilibrium state. However, there is still a lack of explorations of the physical mechanisms behind the whole rich ice front morphology. One may ask: How does the ice front morphology change when the system is tilted? What are the hydrodynamical mechanisms that account for the extent and the complex ice front morphology?  

In this work, by combining experiments, numerical simulations, and theoretical modeling, we aim to systematically explore the freshwater solidification and its coupling with turbulent convective flows to understand the complex behaviors of the ice front morphology at varying system inclination angle, $\beta$ (unit: deg).
The experiments are conducted in a classical Rayleigh-B\'enard (RB) convection system \cite{sig94,bod00,ahlers2009heat,lohse2010small,chi12} (a fluid layer confined between a cold top plate, with temperature, $T_t$, and a hot bottom plate, with temperature, $T_b$), with a quasi two-dimensional rectangular shape (aspect ratios  $L_x/H=1$ and $L_z/H=1/4$, with $H=24~cm$ in experiments). The working fluid is deionized ultrapure water (Prandtl number Pr $\approx 11$). 
The simulations are performed by means of the {\sc Ch4-project} code \cite{calzavarini2019eulerian}, which adopts a Lattice-Boltzmann algorithm for the description of fluid and temperature dynamics, and an enthalpy method for the ice evolution (details see the
Supplemental Material \cite{wang_suppl}) \cite{succi2001lattice,huber2008lattice,Esfahani2018Basal,chen2017a,Moritz2019An,wang2020growth}. Since the water thermal expansion coefficient inverses at $T_\text{c}$ ($\approx 4^\circ$C), here we use the non-monotonous relationship of density with temperature for water near $T_\text{c}$ \cite{Gebhart1977A}, $\rho(T)=\rho_c(1-\alpha^*|T-T_\text{c}|^q )$,
with $\rho_c = 999.972~kg/m^3$ the maximum density  $T_c \approx 4^\circ$C, $\alpha^* = 9.30\times10^{-6} (K^{-q})$, and $q=1.895$. Both in experiments and simulations $T_t$ is fixed at $-10^\circ$C. In the simulations we neglect the microscopic physics leading to kinetic undercooling, Gibbs-Thomson effect and the anisotropic growth/melting \cite{dash2006physics}. 
\begin{figure}[!ht]
	\centering
	\includegraphics[width = 1\textwidth]{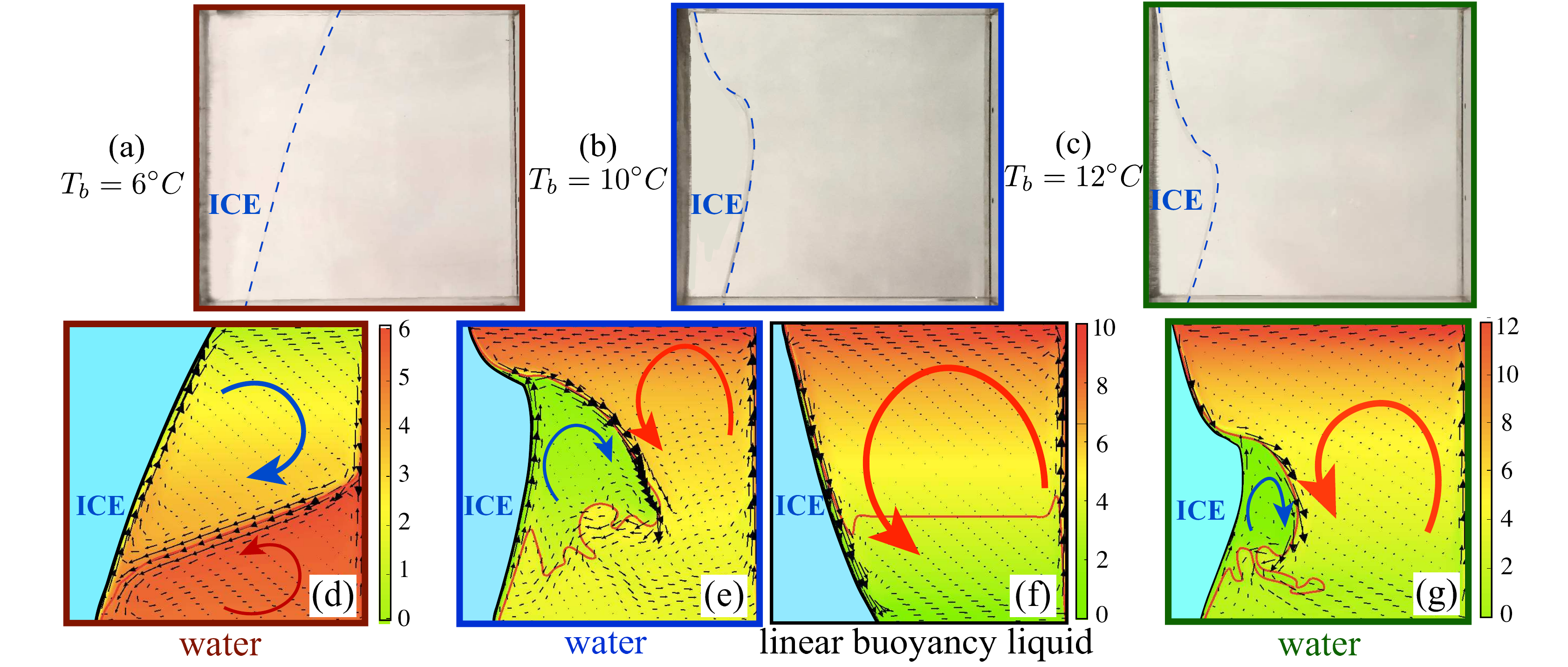}
	\vspace{-0.2cm}
	\caption{Comparison of the ice morphology at the equilibrium state in VC from experiments (a, b, c), simulations with (d, e, g) and without (f) considering the density anomaly. 
	The heating condition is: $T_b = 6^\circ$C (a, d), $10^\circ$C (b, e, f), $12^\circ$C (c, g). (d-g) Temperature fields with color, 0$^\circ$C (black line) and 4$^\circ$C (red line) isotherms, and velocity vectors (black arrows). The red and blue arrows in panels (d-g) show the rotating convective rolls.}
	\label{FIG1} 
\end{figure}
In this study, we monitor the local and the global ice thickness, denoted respectively $h_i(x,t)$ and $h_i(t)$ and expressed in units of the cell height $H$ ($h_i =1$ means full solidification). We consider that the equilibrium is reached when the standard deviation of $h_i(t)$ over a time-window of about 8 minutes is less than 0.5\%. 
We first show the comparison of the ice front morphology at the equilibrium state under different heating conditions, $T_b$, between experiments and simulations, under vertical convection (VC, solidification from left with $\beta=90$). 
Fig.~\ref{FIG1}(a, b, c) are the experimental results with $T_b=6^\circ$C, $10^\circ$C and $12^\circ$C, respectively (more results see the Supplemental Material \cite{wang_suppl}). The corresponding simulation results with properly considering the density anomaly are shown in Fig.~\ref{FIG1}(d, e, g), representing a good agreement with the experimental measurements. This indicates that the simulation indeed can capture the correct behavior of the system. The ice front morphology displays a drastic change as $T_b$ increases. This is due to the competition of two convective rolls originating from the density anomaly, i.e., the one originating from the hot plumes detaching from the hot plate (red arrow in Fig.~\ref{FIG1}(d, e, g)) and the other from the cold upward convective current along the ice front (blue arrow in Fig.~\ref{FIG1}(d, e, g)). The strength of the rolls can be adjusted by changing $T_b$. When $T_b=6^\circ$C (Fig.~\ref{FIG1}(a)), the whole ice is shielded by the upward convective current, and thus the ice front is flat but only tilted. As $T_b$ increases to $10^\circ$C, the ice front becomes highly uneven and is thicker when protected by the upward convective current, thinner at the bottom, and thinnest at the top where the impingement of hot plumes increases the local heat transfer.
Further, when $T_b$ is even higher (Fig.~\ref{FIG1}(c, g)), the anticlockwise roll is much stronger, so it is able to intensively penetrate the clockwise roll and finally affects the ice front, resulting in thinner averaged ice thickness but with a similar shape to that of $T_b =10^\circ$C.

In the simulations, the flow structures and ice front profile are highly different when neglecting the density anomaly (i.e., the density is a linear function of the temperature, see Fig.~\ref{FIG1}(f)): there is only one convective roll, and the ice front is flat and thinner at the top and thicker at the bottom. The clear distinction between the simulation with (Fig.~\ref{FIG1}(e)) and without (Fig.~\ref{FIG1}(f)) considering the density anomaly indicates that this property is crucial to properly describe the ice formation in presence of natural convection. 
\begin{figure}[!ht]
	\centering
	\includegraphics[width = 0.7\textwidth]{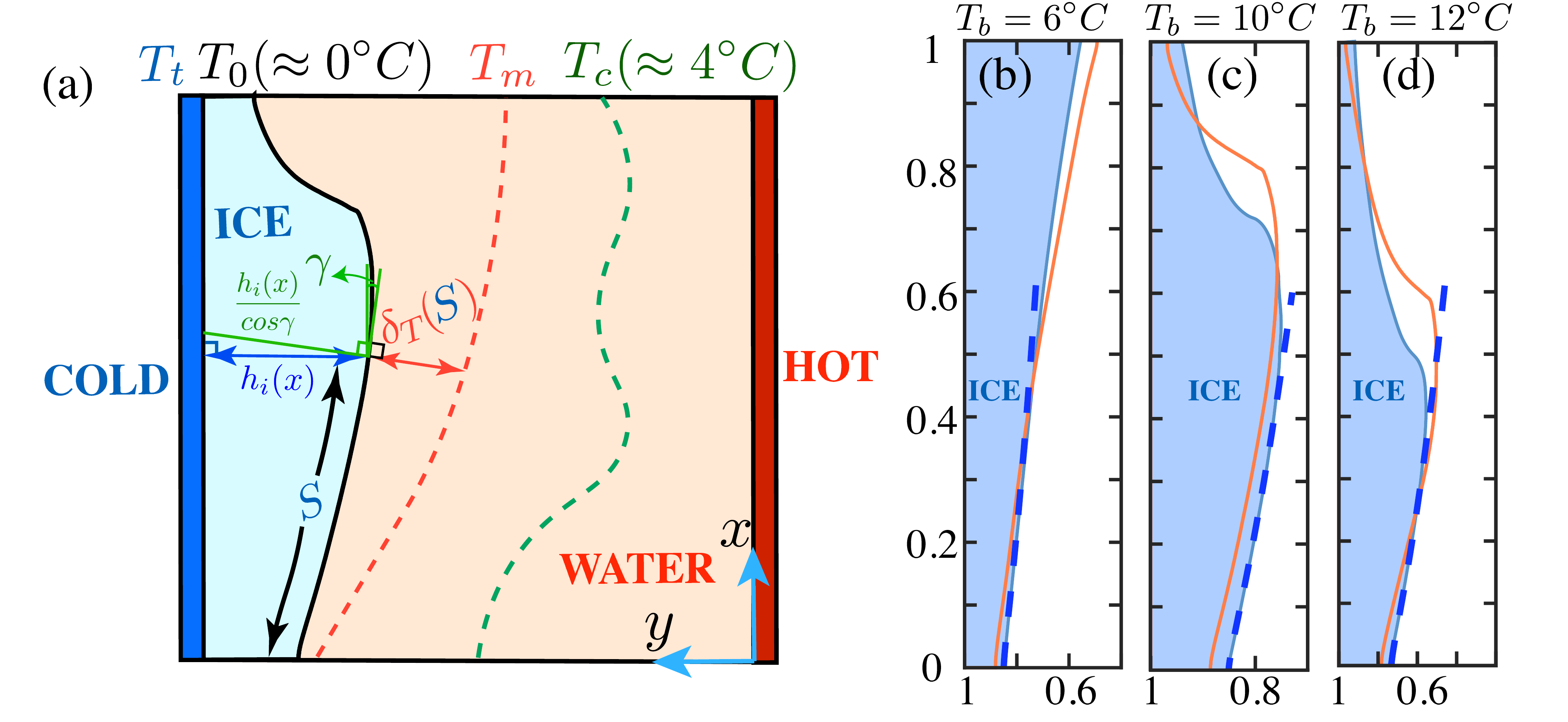}
	\vspace{-0.2cm}
	\caption{Boundary layer model to explain the ice front morphology in the VC case. (a) Sketch of the model: the ice front at $T_0 = 0^\circ$C (black line), 
	the $T_m$ isotherm which is the outer boundary of the thermal boundary (red dashed line), 
	the $T_c$ isotherm (green dashed line). The angle between the tangential direction of the ice front and the x-direction is $\gamma$; the thickness normal to the ice front (the green thick line in the ice) is $\frac{h_i(x)}{cos\gamma}$ (with $cos\gamma = \frac{dS}{dx}$). (b-d) Comparison of ice front morphology among experiments (shaded area), simulations (line) and the model (dashed line) for different $T_b$.
}
		
	\label{blmodel} 
\end{figure}
\begin{figure*}[t!h]
	\centering
	\includegraphics[width = 1\textwidth]{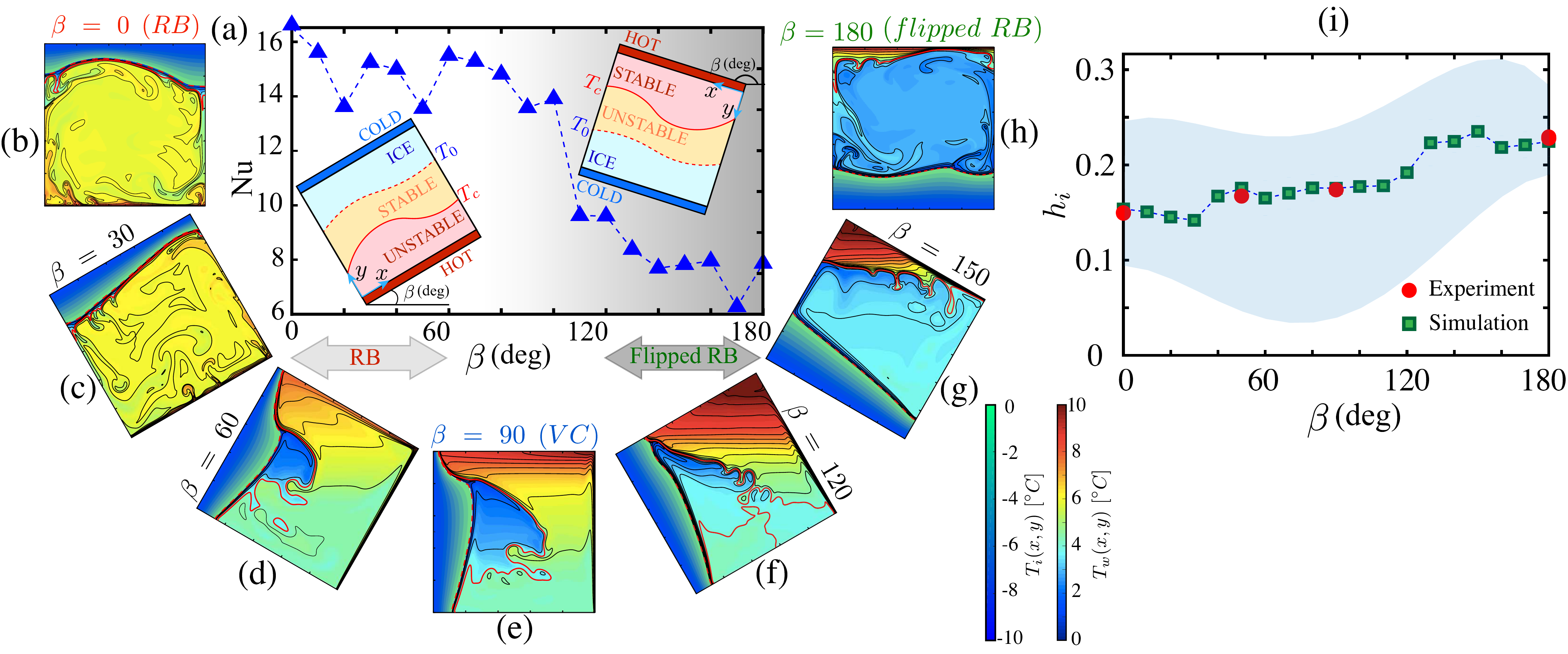}
	\vspace{-0.2cm}
	\caption{(a) Nu as a function of $\beta$.
	Insets: sketches of the arrangement of stably- and unstably-stratified layers of RB regime, $\beta < 90$ (left) and flipped RB, $\beta > 90$(right). 
Error bars (estimated based on the temporal time series of Nu in one simulation) are smaller than the symbols. 
The way to measure the systematic error (i.e., ensemble average, qualitatively reflected by the scatter of the data) is to perform multiple and independent simulations, which is numerically expensive and therefore is not considered here.
	(b-h) Instantaneous temperature field in ice, $T_i(x, y)$, and water, $T_w(x, y)$, and isotherms (black thin lines) at the equilibrium state for different angles $\beta = [0, 180]$.
	The red dashed line and red thick line in panels (b-h) are 0$^\circ$C and 4$^\circ$C isotherms, respectively. (i) Global ice thickness $h_i$ as a function of $\beta$. The shaded area shows the spatial variation of $h_i(x)$ in the simulations. The parameters are $T_b=10^\circ$C and $T_t=-10^\circ$C for all cases.}
	\label{FIG3} 
\end{figure*}

To better understand the morphology of the ice front at its steady state, we introduce a simple model based on the idea that a developing thermal boundary layer forms along the ice-water interface. We note that when the ice has ceased to grow there exists a local balance between the heat flux across the ice and the one across the external boundary layer. The intensity of such fluxes can be estimated by considering the thermal difference across the ice and the boundary layer and the specific geometry of the problem, as shown in the sketch Fig.~\ref{blmodel}(a)). We introduce a curvilinear coordinate, $S$, measuring the length of the ice front starting from its boundary point at the $x=0$ position, and that is linked to the local ice thickness by the arc-length formula $S(x) = \int_{0}^{x} \sqrt{1+(\tfrac{d (h_i(\xi))}{d\xi})^2}\, d\xi$. We now express heat flux balance in the direction normal to the ice surface as,
\begin{equation}
k_i\frac{(T_0-T_t)}{h_i(x) \frac{dS(x)}{dx} }  = k_w \frac{(T_m-T_0)}{\delta_T(S(x))}, 
\label{eq:1}
\end{equation}
where $k_w$ and $k_i$ are the thermal conductivity of water and ice, $(T_0-T_t)$ is the temperature difference in the ice and $h_i(x) / \tfrac{dS(x)}{dx}$ is the ice thickness in the direction normal to the ice front. The estimation of the heat-flux in the water involves making further assumptions. First, we consider that the heat transport in the boundary layer in the direction normal to the ice front is purely conductive, and this is justified by the fact that for $Pr \gg 1$ the thermal boundary layer is nested inside the viscous one \cite{sun2008experimental}. Second, we assume that the thermal difference across the boundary layer of the fluid layer is half of the one in the adjacent recirculation region (denoted with the blue arrow in Fig.~\ref{FIG1}(d, e, g)), i.e. $(T_m-T_0)$, with $T_m = (T_0+T_c)/2$. Third, the boundary layer thickness $\delta_T(S)$ is assumed to vary along the ice front, as in a developing vertical thermal boundary layer, with a dependence that we take to be $\delta_T(S) = C_1 (S  + C_2)^{1/4}$ \cite{bejan2013convection,white2006viscous,shishkina2016momentum}. The latter expression has two dimensionless parameters 
$C_1 = c \cdot (g [1-\rho(T_m)/\rho_c]/(\nu \kappa))^{1/4}$ with $c$ the proportional constant ($c \approx 5~ \text{m}^{3/4}$, $c$ has the unit m$^{3/4}$ to make $C_1$ dimensionless) \cite{bejan2013convection,white2006viscous,shishkina2016momentum}, 
and the offset $C_2$ because of non-zero boundary layer thickness at $x=0$, with $C_2 =(h_i(0)[k_w(T_m-T_0)]/[k_i(T_0-T_t)]/C_1)^{1/4} $, with the boundary ice thickness $h_i(0)$ as an input from the simulation results. Since the sidewall is adiabatic, $\frac{dh_i}{dx}|_{x = 0} = 0$ is an extra known condition.
With the above choices Eq.(\ref{eq:1}) becomes an integrodifferential equation that can be solved numerically for the local ice thickness $h_i(x)$.
Fig.\ref{blmodel}(b-d) show the comparison of the model prediction with experiments and simulations. A good qualitative agreement is reached in the region where the upward convective current takes place. The disagreement in the upper part of the ice front is expected as the boundary layer no longer develops in that region due to the downward warmer convective current (denoted with the red arrow in Fig.~\ref{FIG1}(d, e, g)). Although the exact spatial dependence of $\delta_T$ along the ice interface is not known, the present model put forward a robust physical mechanism for the shaping of the ice in the bottom part of the convection cell.\\

Next, we perform systematical numerical simulations to explore how $\beta$ affects the extent and morphology of the ice front, with 0$ \leq \beta \leq180$ (with coordinate system attached to the cell). We limit our study to $T_b=10^\circ$C, and the results are easy to extend to other situations.

First, we calculate the heat transfer rate, which when expressed dimensionlessly is the global Nusselt number (Fig.~\ref{FIG3}(a)), Nu$ =(\left\langle u_yT\right\rangle -\kappa \partial_y \left\langle T\right\rangle )/(\kappa \Delta T/H)$,
  where $\langle...\rangle$ represents an average over time and the whole cell volume, $H$ is the system height. It is noteworthy that there is convection both at $\beta=0$ (heating from below) and $\beta = 180$ (heating from above).
Remarkably, as $\beta=180$, the fluid is unstably stratified in the temperature interval between $T_0$ and $T_c$ which accounts for creating the convection. The latter feature is specific to water and does not occur in other systems with working fluid's density increasing with temperature. 
\begin{figure*}[t!h]
	\centering
	\includegraphics[width = 1\textwidth]{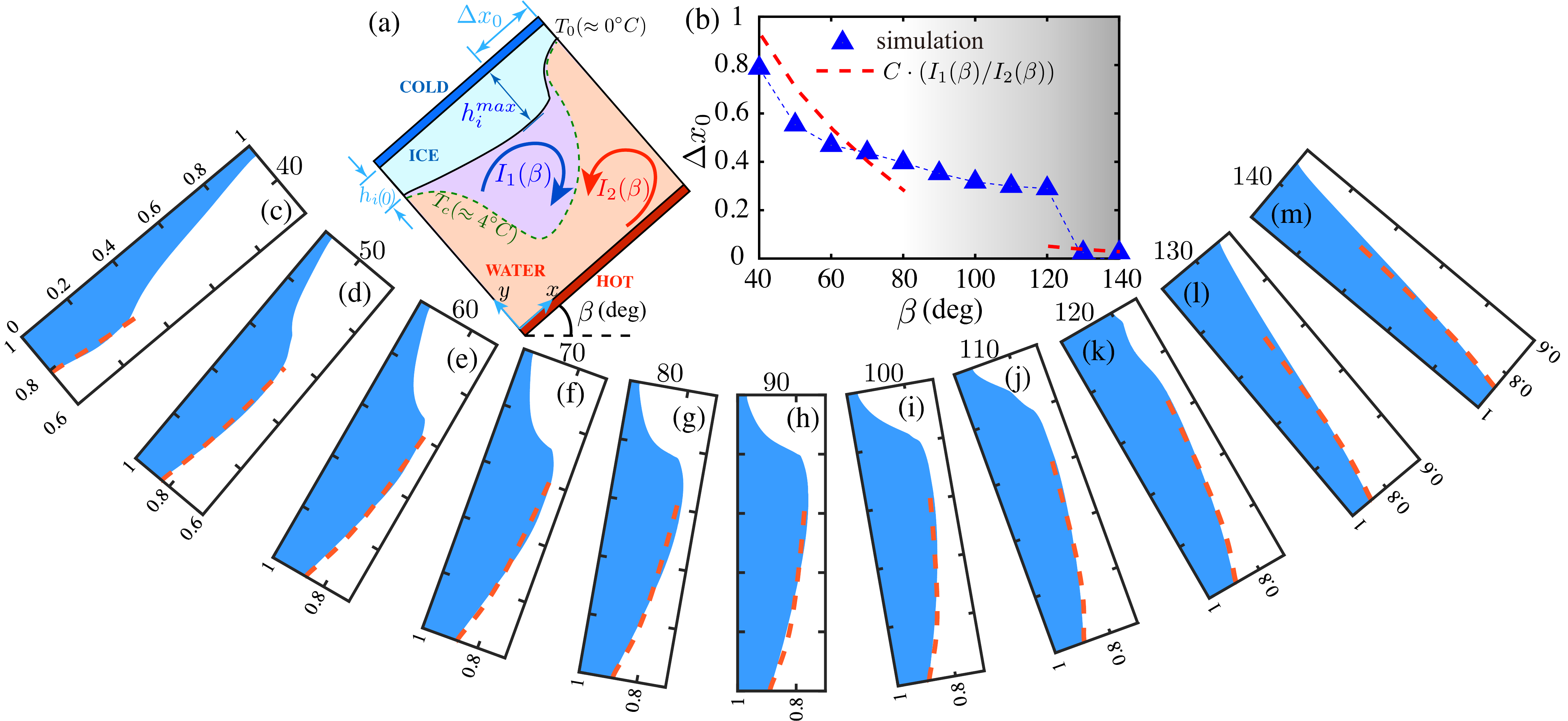}
	\vspace{-0.2cm}
	\caption{(a) Sketch of the buoyancy-intensity model to explain the position of the maximum ice thickness. The buoyancy intensity of the clockwise and anticlockwise convective rolls are
		$I_1(\beta)$ and 
		$I_2(\beta)$, respectively. $\Delta x_0$ locates the position of the maximum of the $h_i(x)$, which is dimensionless normalized by $H$. 
		$h_i(0)$, an input parameter of the boundary layer model, is roughly constant and around $h_i$. $h_i(0)$ vs. $\beta$ is reported in the Supplemental Material \cite{wang_suppl}. (b) $\Delta x_0$ as a function of $\beta$and its comparison with model predictions (adjustable constant C=0.08).
		(c-m) Comparison of the ice front morphology between simulations (shaded area) and the model (dashed line) under different $\beta$ near $90$.}
	\label{FIG4} 
\end{figure*}
Fig.~\ref{FIG3}(b-h) show the temperature field with different $\beta$ from simulations (more results see the
Supplemental Material \cite{wang_suppl}). The inclination results in different levels of thermal stratification, which induces huge modifications to the ice front morphology. When $\beta = 0$, 
there is a stably-stratified layer (from $T_0$ to $T_c$) on top of the unstably-stratified layer (from $T_c$ to $T_b$). As the cell is tilted with a small $\beta$, the convection is strong enough to squeeze the stably-stratified layer to be closely attached to the ice front, so the ice morphology is influenced by a single-roll convective flow (see Fig.~\ref{FIG3}(b, c)). As $\beta$ increases towards $90$, the inclination of the temperature gradient with respect to the gravity is strong enough to break down the stratification and thus the stably-stratified layer is set into motion in the form of a clockwise convective roll (an upward cold water current) which competes with the initially anticlockwise roll (downward warm water current). The shielding effect is prominent because the ice thickness reaches a local maximum, and the hot plumes impacting on the top part of the ice induces a local minimum of the ice thickness, based on these two kinds of effects, the ice front presents an inflection point during the transition from the thickest to the thinnest part, which has also been reported in Fig.~\ref{FIG1}(b, c). The flow motion of the original stably-stratified layer intensifies as $\beta$ increases for $\beta < 90$. Beyond $\beta=90$, the stratification configuration flips over. On the whole, the intensity of convection is higher at $0 < \beta < 90$ than that of $90 < \beta < 180$. This is to be connected to the different thermal difference across the respective unstably stratified layers which are $(T_c-T_0)\approx 4$K in the former and $(T_b-T_c)\approx6$K in the latter case.  This change in driving strength also accounts for the observed results oF Nu (Fig. 3(a)).

To account for the influence of $\beta$ in a quantitative way, we now calculate stationary global ice thickness ratio, $h_i = \langle h_i(x,y)\rangle_{x,t}$, where $\langle...\rangle_{x,t}$ represents an average over time and x-axis direction.
Fig.~\ref{FIG3}(i) reports $h_i$, as a function of $\beta$.  $h_i$ shows an increasing trend as $\beta$ increases because the heat transfers less efficiently for large $\beta$ (i.e., flipped RB regime). The results from experiments (red circles) and simulations (green squares) agree well with each other. It is noteworthy that the ice front is highly variable in space as a result of different coupling with turbulent flow structures. Here, the spatial fluctuations of the ice thickness are represented by the local maximum and minimum of $h_i$, which is highlighted by shaded area.

Coming back to cases when $\beta$ is around 90, we can observe a peculiar yet robust form of ice front morphology (which can be observed in the range of $\beta=40\sim140$). 
In fact, the aforementioned boundary layer model can be extended to the tilted system, by modifying $C_1$, with 
$C_1 = ((g_x (1-\rho(T_m)/\rho_c) - g_y \mathcal{H}[\beta-90^{\circ}] (1-\rho(T_0)/\rho_c))/(\nu \kappa))^{1/4}$,
where $\mathcal{H}$ is Heaviside step function, $g_x = g sin\beta$, $g_y = g cos\beta$. The first term in {$C_1$} results from the inclination effect, the second part originates from the inherent buoyancy contribution induced by the density difference, and is present only for inclinations larger than $90^{\circ}$.
Fig.~\ref{FIG4}(c-m) show the comparison of the ice front morphology between simulations (shaded area) and model prediction (dashed line). As shown, the model can qualitatively capture behaviors of the ice front at the inception of the thermal boundary.
Note when $\beta < 40 $ and $\beta >140$, 
the boundary layer attaching to the ice front is disturbed because of the plume impacting under intensive interactions of the stably- and unstably-stratified layers, so the model cannot be utilized. 
A second limitation of the model is that the adopted expression for $\delta_T(S)$ is based on VC \cite{bejan2013convection} but it does not involve possible dependencies on the cell inclination. Nevertheless, it is remarkable that the model already performs well in a wide range of $\beta$. Further studies are needed to improve or quantitatively refine the model.

Another feature of the ice front is the position, $\Delta x_0$, where $h_i(x)$ reaches maximum (symbols in Fig.~\ref{FIG4}(b)) and also implies the extent of hot plume impacting effect. 
As discussed before, the local maximum of $h_i$ originates from the competition of the buoyancy intensity between two counterrotating rolls. The buoyancy intensity of the cold (clockwise) roll (blue arrow in Fig.~\ref{FIG4}(a)), $I_1(\beta)$, and of the warm (anticlockwise) roll (red arrow in Fig.~\ref{FIG4}(a)), $I_2(\beta)$, can be approximately evaluated with
\begin{equation}
\begin{split}
&I_1(\beta) =g_x (1- \tfrac{\rho(T_m)}{\rho_c}) - g_y \mathcal{H}[\beta-90^{\circ}] (1-\tfrac{\rho(T_0)}{\rho_c});\\
&	I_2(\beta) =g_x (1-\tfrac{\rho(T_{m2})}{\rho_c}) + g_y \mathcal{H}[90^{\circ}-\beta] (1-\tfrac{\rho(T_b)}{\rho_c}).\\
\end{split}
\label{buoyancyintensity}
\end{equation}
where the mean temperature $T_m = (T_0+T_c)/2$ and $T_{m2} = (T_b + T_c)/2$. 
 Eqn.~(\ref{buoyancyintensity}) holds when $\beta \neq 90$ (details see the Supplemental Material \cite{wang_suppl}). The intensity ratio, $(I_1(\beta)/I_2(\beta))$, captures the trend of $\Delta x_0$ as a function of $\beta$, at least for $10<|\beta - 90|<50$. A quantitative agreement is obtained by an adjustment multiplicative factor $C\simeq 0.08$ (red dashed line in Fig.~\ref{FIG4}(b)).
This heuristic model provides further evidence that the competition of two convective rolls accounts for the form of ice morphology.

To summarize, we found that the existence of a cold upward convective current, due to the density anomaly of water, accounts for the ice shaping. 
We provide physical understanding on the main features of the ice morphology in a wide range of system inclinations. The present exploration offers deeper insight into comprehending the liquid-solid interface morphology induced by the coupling between phase-transition and natural convection with possible applications in geophysical and climate sciences.

\begin{acknowledgments}
This work was supported by Natural Science Foundation of China under grant nos 11988102, 91852202 and 11861131005.
\end{acknowledgments}


\vspace{-0.7cm}
\clearpage
\onecolumngrid
\appendix

\setcounter{equation}{0}
\setcounter{figure}{0}
\setcounter{table}{0}
\setcounter{page}{1}
\makeatletter
\renewcommand{\theequation}{S\arabic{equation}}
\renewcommand{\thefigure}{S\arabic{figure}}

	\newpage
{	\Large{\textbf{Supplementary Information for: \\Ice front shaping by upward convective current}
}}

	\section{Experimental Methods}
	
\begin{figure*}[!htb]
		\centering
		\includegraphics[width=0.8\linewidth]{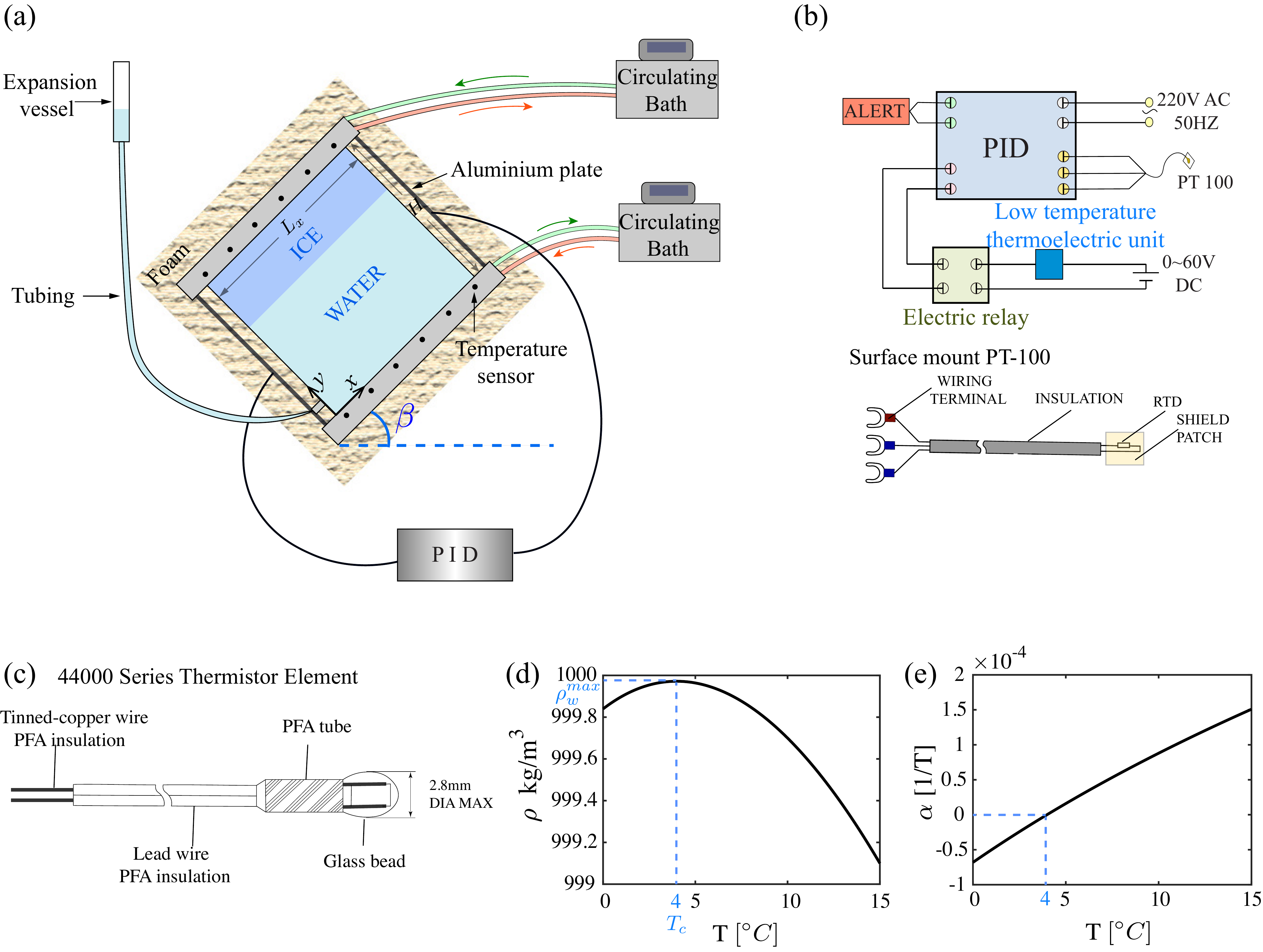}
		\caption{(a) Sketch of the experimental system for Rayleigh-Bénard convection coupled with solidification of fresh water under different inclination angles. 
			(b)The PID (Proportional-Integral-Derivative) controller and the temperature sensor used in the PID system. (c) The sketch of resistance thermistor, 44000 series thermistor element, which is used to measure the top and bottom plates temperature. (d)  The nonmonotonic relationship of the water density with temperature for cold water near $T_c$ from Ref. \cite{Gebhart1977A}. (e) the thermal expansion coefficient of water as a function of the temperature.}
		\label{SM_expsetup} 
	\end{figure*}

	The experiments of freshwater solidification in a convective cell with different inclination angles (here we use the coordinate system attached to the cell) are conducted in a classical Rayleigh-B\'enard (RB) convection system (see Fig.~\ref{SM_expsetup}).  Fig.~\ref{SM_expsetup}(a) shows the sketch of the experimental cell. The experimental cell consists of a top cooling plate (with temperature $T_t$), a heating bottom plate (with temperature $T_b$), and plexiglas sidewalls with height H = 240 mm (length $L_x =240$ mm and width $L_z =60$ mm, i.e., aspect ratio $\Gamma = L_x/H = $1.0). The working fluid, which is confined between the top and bottom plates, is deionized and ultrapure water, and the water density and thermal expansion coefficient around the density peak temperature, $T_c=$4$^\circ$C, are shown in Fig.~\ref{SM_expsetup}(d) and (e). Before conducting any experiments, water is boiled twice to be degassed. The temperature of both the top and bottom plates are well controlled at constant temperatures ($T_t < 0^\circ$C, and $T_b > 0^\circ$C) by the circulating bath (PolyScience PP15R-40), with the temperature fluctuations less than $\pm0.2$K. 
	Silicone O-rings are attached in between the top plate and the sidewall, and also in between the sidewall and the bottom plate, to seal the cell. In order to compensate the volume change during the solidification of freshwater, an expansion vessel is connected to the experimental cell through a rubber tube, which is open to the atmosphere so that the pressure of the experimental cell remains unchanged. Six resistance thermistors (44000 series thermistor element, see the sketch in Fig.~\ref{SM_expsetup}(c)) are embedded into the top and bottom plates, respectively (see the sketch in Fig.~\ref{SM_expsetup}(a), the black dots on the top and bottom plates show the placement of the thermistors). There are two steps to avoid the heat exchange between the experimental cell and the surrounding environment, 1) the experimental cell is wrapped in a sandwich structure: insulation foam, aluminum plate (helping to support and acting as the temperature measuring spot of the temperature sensor, PT 100, of the Proportional-Integral-Derivative (PID) system), and insulation foam; 2) a PID (Proportional-Integral-Derivative) controller is used to control the temperature of the surrounding environment outside the experimental cell (see Fig.~\ref{SM_expsetup}(b)).
	
	During the experiments and numerical simulations, in order to focus on the effect of inclination, we have limited the cold plate to a constant undercooling temperature of $T_t = -10^\circ$C, which is a typical value in winter. The effects of changing the undercooling temperature, $T_t$, are qualitatively predictable and are not expected to change the occurrence of different forms of the ice front morphology upon increasing the inclination angle, $\beta$, as well as its physical explanation (by the boundary layer theory based model). The only effects that $T_t$ will bring are that, 1) the ice thickness at the equilibrium state of the system is thicker (thinner) with decreasing (increasing) $T_t$; 2) the corresponding ice front morphology is similar in the shape compared with that of the current study but will be different in the extension and the local curvature. 
	
	We have conducted several typical sets of experiments (we only perform the typical cases of the experiments and did not conduct every experiment corresponding to the simulations because the experiments are time consuming): 1) for investigation of the effects of different heating condition, we have performed $T_b = 6^\circ$C,  $10^\circ$C and $12^\circ$C, at the system inclination angle of $\beta=90$; 2) for the investigation of the influence of the cell inclination, we have done several different inclination angles, $\beta$ (unit: deg), i.e., $\beta=0$, 50, 90, 180, at the hot plate temperature $T_b =10^\circ$C;

	In the experiments, the ice forms on the cold plate and grows in thickness until the system reaches a statistical equilibrium state. The statistical equilibrium state is reached when the standard deviation of the ice thickness time series is less than 0.5\%. 
	
	Changing the heating condition, $T_b$ can adjust the intensity of the two counterrotating convective rolls. Recall that the water density reaches the maximum, $\rho_c$, at $T_c$ (which is approximately $4^\circ$). Then depending on the value of the hot plate temperature, $T_b$, the flow configuration can be classified into two regimes: 
	
	1) When $4^\circ \text{C}< T_b \leq 8^\circ \text{C}$, the anticlockwise convective roll (with the temperature ranging from $T_c$ to $T_b$ and the corresponding thermal driving $T_b-T_c \leq 4$K) has weaker intensity than that of the clockwise (with the temperature ranging from $T_0$ to $T_c$ and the corresponding thermal driving $T_c-T_0 \approx 4$K). In this regime, the whole ice front is shielded by the colder clockwise roll and is away from the influence of the hot plumes from the hot plate. So the ice front is flat with a tilting angle.
	
	2) When $T_b > 8^\circ \text{C}$, the anticlockwise convective roll (with the temperature ranging from $T_c$ to $T_b$ and the corresponding thermal driving $T_b-T_c > 4$K) has stronger intensity than that of the clockwise (with the temperature ranging from $T_0$ to $T_c$ and the corresponding thermal driving $T_c-T_0 \approx 4$K). So the anticlockwise roll penetrates the clockwise roll and finally affects the ice front morphology, which represents thinner in thickness with higher $T_b$ but similar shape.
	
\begin{figure*}[!htb]
		\centering
		\includegraphics[width=0.6\linewidth]{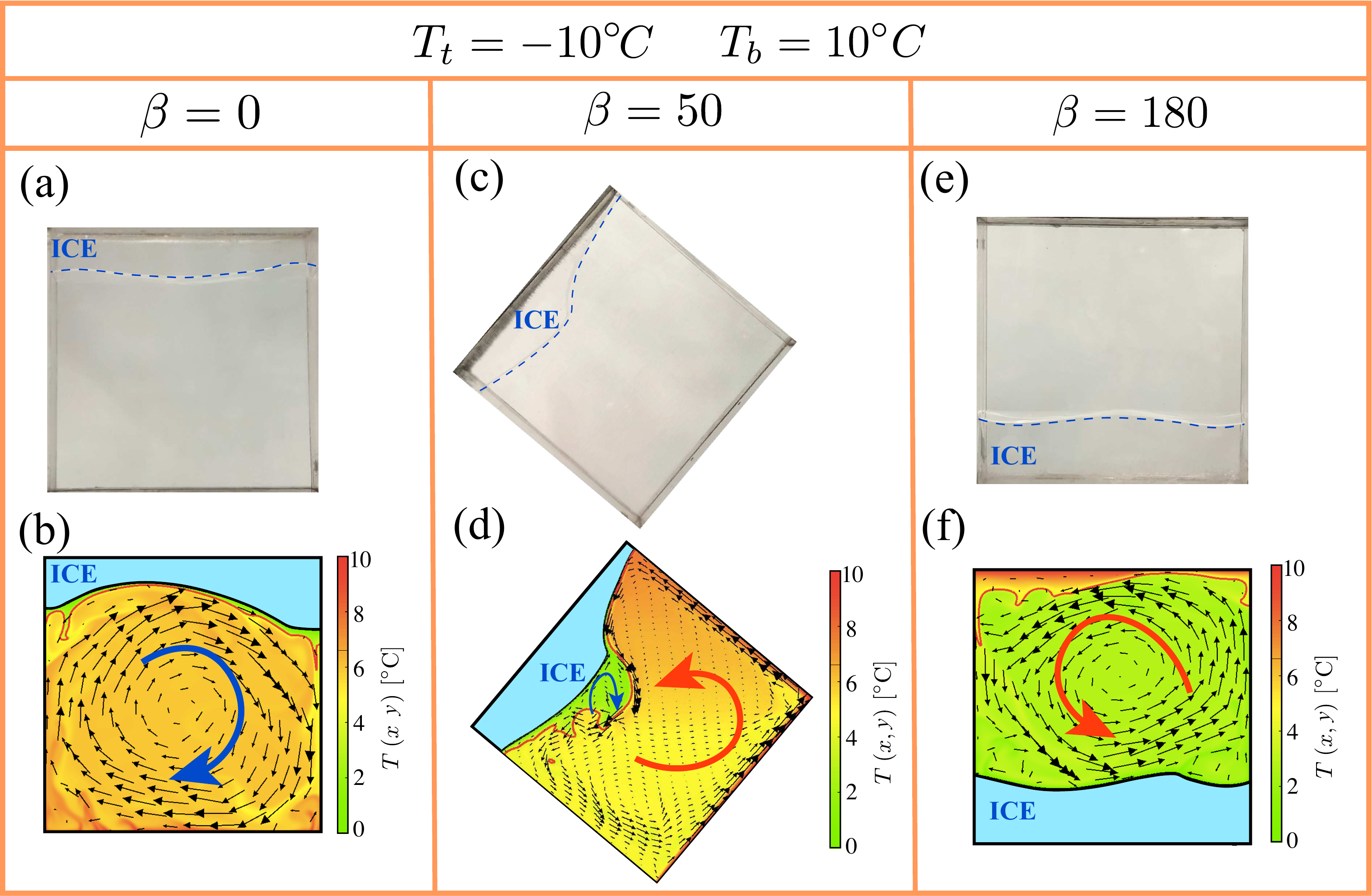}
		\caption{More results of the experiments ((a), (c), (e)) compared with the simulations (with properly considering the water density anomaly) ((b), (d), (f)). Parameter chosen: (a) and (b): $T_b = 10^\circ$C and $\beta=0$; (c) and (d): $T_b = 10^\circ$C and $\beta=50$; (e) and (f): $T_b = 10^\circ$C and $\beta=180$. The unit for $\beta$ is deg. The blue dashed line in (a), (c), and (e) shows the ice front position which is guide to the eye. (b), (d), and (f) are the temperature field overlapped with isotherms as well as the velocity vectors. The ice front is represented by the black thick line and the $T_c$-isotherm is represented by the red thick line. The blue shaded area is the ice layer.}
		\label{SM_moreexppic} 
	\end{figure*}

	With the knowledge of these, we later limited our investigation to the case with $T_b =10^\circ$C and performed more experimental investigation under different inclination angles (see Fig.~\ref{SM_moreexppic}).

	Fig.~\ref{SM_moreexppic} reports more experimental results with the parameter $T_b=10^\circ$C and $T_t=-10^\circ$C, with inclination angle, $\beta=0$(panels (a) and (b)), $\beta=50$(panels (c) and (d)), and $\beta=180$(panels (e) and (f)). Fig.~\ref{SM_moreexppic}(a), (c), and (e) are from experiments, and Fig.~\ref{SM_moreexppic}(b), (d), and (f) are from the corresponding simulations. 

	From Fig.~\ref{SM_moreexppic}, we can conclude that the results from the experiments and numerical simulations under different inclination angles agree well with each other.
	
	{\color{black}
		
		\section{Numerical Methods}
		
		The simulations are performed by means of the \textsc{CH4-PROJECT} code \cite{calzavarini2019eulerian}, which adopts a Lattice-Boltzmann algorithm for the description of fluid and temperature dynamics, and an enthalpy method for the ice evolution. The method has been validated against experiments  \cite{wang2020growth}, while the code has been intensively tested in \cite{Esfahani2018Basal}.

		In the simulation, we use the Boussinesq approximation, which means that the density is regarded as a constant value except for that in the buoyancy term in the momentum equation. The non-monotonic relationship of the water density as a function of the temperature is $\rho_\text{w} = \rho_c(1-\alpha^*|T_b-4|^q )$ \cite{Gebhart1977A}, 
		with $\rho_c = 999.972~kg/m^3$ the maximum density corresponding to $T_c \approx 4^\circ$C, $\alpha^* = 9.30\times10^{-6} (K^{-q})$, and $q=1.895$. All the physical properties of water and ice phase, except for $\rho_w$ in the buoyancy term are evaluated at the mean temperatures in each phase which are $(T_\text{b}+0)/2$ and $(T_\text{t}+0)/2$, respectively. In the simulations we neglect the microscopic physics leading to kinetic undercooling, Gibbs-Thomson effect and the anisotropic growth/melting \cite{dash2006physics}. Furthermore, we assume the ice and water density remain the same, i.e., $\rho_I = \rho_w$, to satisfy the incompressibility of the flow in the simulation. 
		The relevant equations that govern the fluid flow and the boundary conditions are reported here below (the energy equation will be presented later),
		\begin{equation}
			\begin{split}
				&\vec{\nabla} \cdot \vec{u} = 0,\\
				& \frac{\partial \vec{u}}{\partial t}  + \vec{u}\cdot \vec{\nabla} \vec{u} = -\frac{\vec{\nabla} p}{\rho_\text{c}} +\nu_w \nabla^2 \vec{u} +\alpha^*g|T-4|^q  \hat e_z,
			\end{split}
			\label{governing_eqn}
		\end{equation}
		where $\vec{u}(x,y,t)$, $p(x,y,t)$ are fluid velocity, pressure, respectively and 
		we denote with $x$ as the horizontal direction and with $y$ the vertical direction; $\nu_w$, $\rho$, $g$ are the kinematic viscosity of water, the density, and the acceleration of gravity, respectively. $\hat e_z$ is the unit vector pointing in the direction opposite to that of gravity.

		\begin{figure*}[t!hb]
			\centering
			\includegraphics[width=0.6\linewidth]{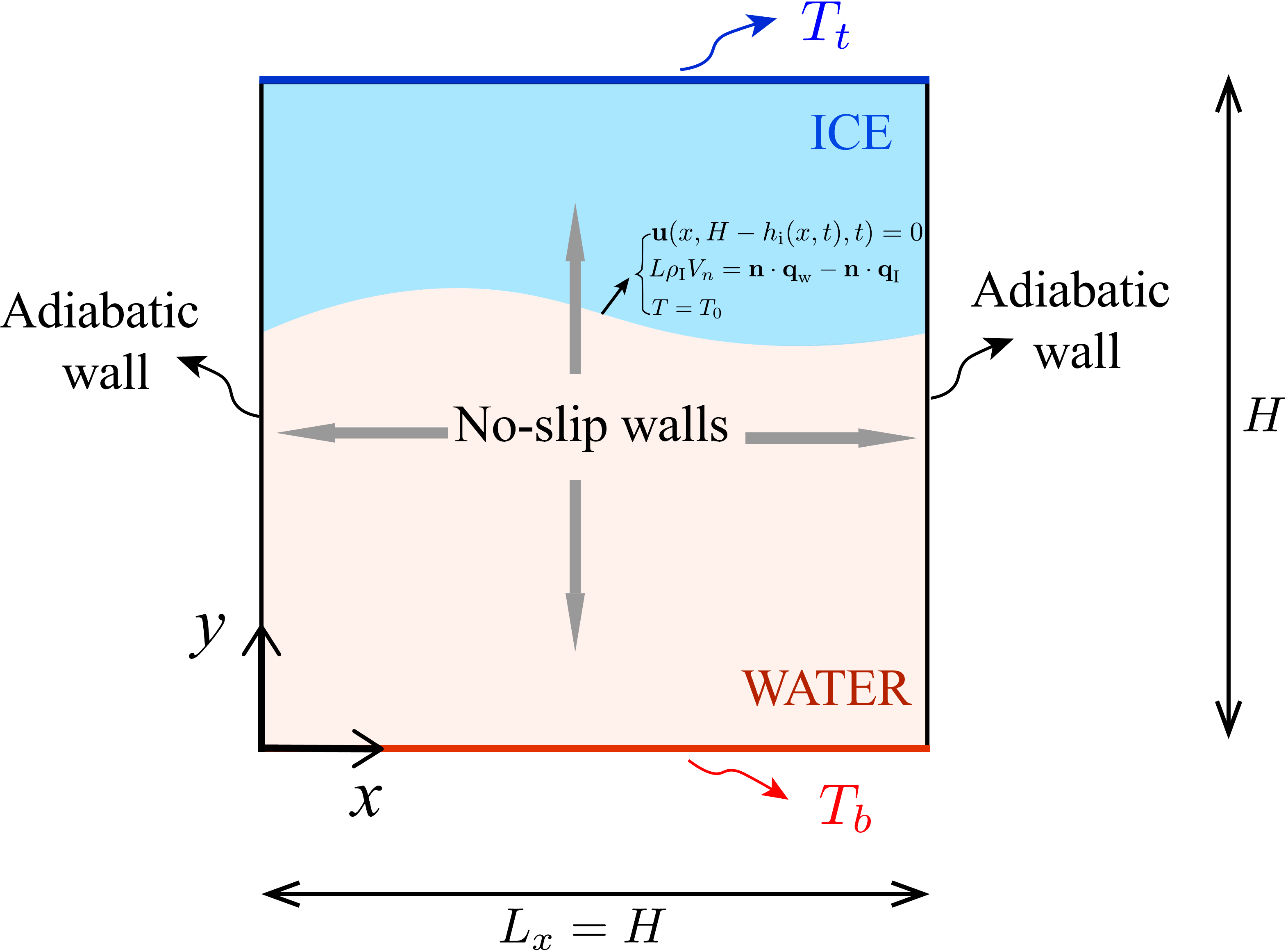}
			\caption{Schematic description of the problem considered. The blue shaded region corresponds to the solid phase (ice) and the red shaded region to the liquid phase (water). $T_0$ is the melting temperature. The top plate is cooled at constant temperature $T_t$, and the bottom plate is heated at the constant temperature $T_b$. The ice-water interface is fixed at $T_0$ and is of no slip condition.The two side walls are of adiabatic conditions.}
			\label{sketch} 
		\end{figure*}	
		
		The boundary conditions corresponding to the governing equations above are isothermal at the top and bottom plates, no-slip at the bottom plate and at the ice-water interface, adiabatic at the lateral boundaries, and no-slip and freezing (namely, Stefan condition \cite{
			bodenschatz2000recent}) at the phase-changing interface (see also figure \ref{sketch}). 
		The boundary conditions read:
		
		\begin{equation}
			\begin{split}
				& T(x,0,t) = T_\text{b},\\
				& T(x,1,t) = T_\text{t},\\
				& \vec{u}(x,0,t) = 0,\\
				& \vec{u}(x,1-h_\text{i}(x,t),t) = 0,\\
				&\frac{\partial {T(x,y,t)}}{\partial x}|_{x=0} = 0,\\
				&\frac{\partial {T(x,y,t)}}{\partial x}|_{x=L_x/H} = 0,\\
				&L \rho_\text{i} V_n = \vec{n} \cdot \vec{q}_\text{w} - \vec{n}\cdot \vec{q}_{\text{i}},
			\end{split}
			\label{governing_eqn_bcs}
		\end{equation}
		where $L$ is the latent heat, $V_n$ is the normal speed of the ice-water interface, and $h_i(x,t)$ the dimensionless ice layer thickness at the position $x$, $\vec{q}$ is the heat flux vector, $\vec{n}$ is a unit normal at the ice-water interface pointing into the liquid. The subscripts I and w refer to the ice and the water, respectively. The heat flux reads $q_\text{i} =  -k_\text{i} \vec{\nabla} T_\text{i}$ and $q_\text{w} =  -k_\text{w} \vec{\nabla} T_\text{w}$.
		
		The boundary condition at the ice-water interface requires particular care due to its time and space dependent character.
		So an useful method is to separate the total enthalpy $h$ into sensible heat and latent heat \cite{Moritz2019An}: 
		
		\begin{equation}
			\text{h} = \left\{
			\begin{split}
				&L\phi_w + \text{C}_\text{pi}T, ~~~~~~~~~~~~~~~~~~~~~~~~~~~when~ T<T_0,\\
				&L\phi_w + \text{C}_\text{pi}T_0, ~~~~~~~~~~~~~~~~~~~~~~~~~~when ~T=T_0,\\
				&L\phi_w + \text{C}_\text{pi}T_0 + \text{C}_\text{pw}(T-T_0), ~~~~~when ~T>T_0.
			\end{split}
			\right.
			\label{enthalpy}
		\end{equation}
		where $T_0$ is the phase change temperature ($T_0=0$), and $\phi_w(x,y,t)$ is the liquid fraction in the system and the relation between $h_i(x,t)$  and $\phi_w(x,y,t)$ is $h_i(x,t) = 1 - \int_{0}^{1} \phi_w(x,y,t)\, dy$. In the ice phase, $\phi_w=0$, and in the water phase, $\phi_w=1$, which leads to an additional source term, $S_1$ (the formulation is shown later), from the latent heat contribution at the ice-water interface in the energy conservation equation.
		
		We use the Lattice Boltzmann method (LBM) which is able to capture the turbulent convective dynamics in the water phase and also describe the phase change process at the ice-water interface. The basic principle and formulation of the method are described in \cite{succi2001lattice}, while the specific applications to phase-change have been presented in \cite{huber2008lattice,Esfahani2018Basal}. It is noteworthy that the key to accurately solve such problems is to recover the diffusion term in the energy conservation equation exactly and, similarly to \cite{chen2017a}, we implement the correction when the investigated domain consists of heterogeneous media which lead to another additional source term, $S_2$, in the energy conservation equation. So the energy equation with consideration of two source terms $S_1$ and $S_2$ reads
		
		\begin{equation}
			\sigma (\rho \text{C}_\text{p})_0 \frac{\partial T}{\partial t} + \vec{ \nabla} \cdot (\sigma (\rho \text{C}_\text{p})_0  T\vec{u}) = \vec{\nabla} \cdot (k \vec{\nabla} T) +S_1+S_2,
			\label{source_term}
		\end{equation}
		where $T(x,y,t)$ is the temperature fields (all temperatures are measured in Celsius), and $k$ is the thermal conductivity. When it is in water phase $k = k_{w}$, $\text{C}_\text{p}= \text{C}_\text{pw}$, and when it is in ice phase $k = k_{i}$, $\text{C}_\text{p}= \text{C}_\text{pi}$. The first source term is $S_1=- L\rho\frac{\partial \phi_w}{\partial t}$ and the second source term $S_2=-\sigma k \vec{\nabla} T \vec{\nabla} \frac{1}{\sigma} - 
		\frac{\rho \text{C}_\text{p}}{\sigma} T \vec{u} \vec{\nabla} \sigma$. Here $\sigma = \frac{\rho \text{C}_\text{p}}{(\rho \text{C}_\text{p})_0}$ is the ratio of heat capacitance (which is variable and depends on the type of phase, i.e., ice or water) and $(\rho \text{C}_\text{p})_0$ is reference heat capacitance which is taken as constant \cite{chen2017a}.
		
		In this study we focus on the morphology of the ice front at the statistical equilibrium state. In order to speed up the process of reaching an equilibrium state, all the simulations are started with a thin layer of flat ice.

		\section{Systematic investigation of various inclination angles}
		
		We have conducted a systematical exploration of how $\beta$ (unit:deg) affects the ice morphology, ranging from $\beta = 0$ to $\beta = 180$ (here we use the coordinate system attached to the cell). Fig.~\ref{SM_systematic_angle} reports the time sequences of the temperature field under different inclination angles, $\beta$ with the parameters $T_t = -10^\circ$C. We focus on the morphology of the ice front at the statistical equilibrium state. In order to speed up the process of reaching an equilibrium state, all the simulations are started with a thin layer of flat ice.   
		
		The boundary conditions are no-slip for the velocity, adiabatic at the sidewalls, and constant temperatures at the top and bottom plates. The initial condition is still fluid at linear temperature profile from $T_0$ to $T_b$. We assume thermophysical properties to be constant except for the density in the buoyancy term. The real water density property near to the density peak temperature, $T_c$, is well described with the equation $\rho(T)=\rho_c(1-\alpha^*|T-T_\text{c}|^q )$,
		with $\rho_c = 999.972~kg/m^3$ the maximum density corresponding to $T_c \approx 4^\circ$C, $\alpha^* = 9.30\times10^{-6} (K^{-q})$, and $q=1.895$
		This form of equation gives the maximum density, $\rho_{c} = 999.972 kg/m^3$, at the density peak temperature $T_c$ \cite{Gebhart1977A}.
		
		\begin{figure*}[!htb]
			\centering
			\includegraphics[width=0.8\linewidth]{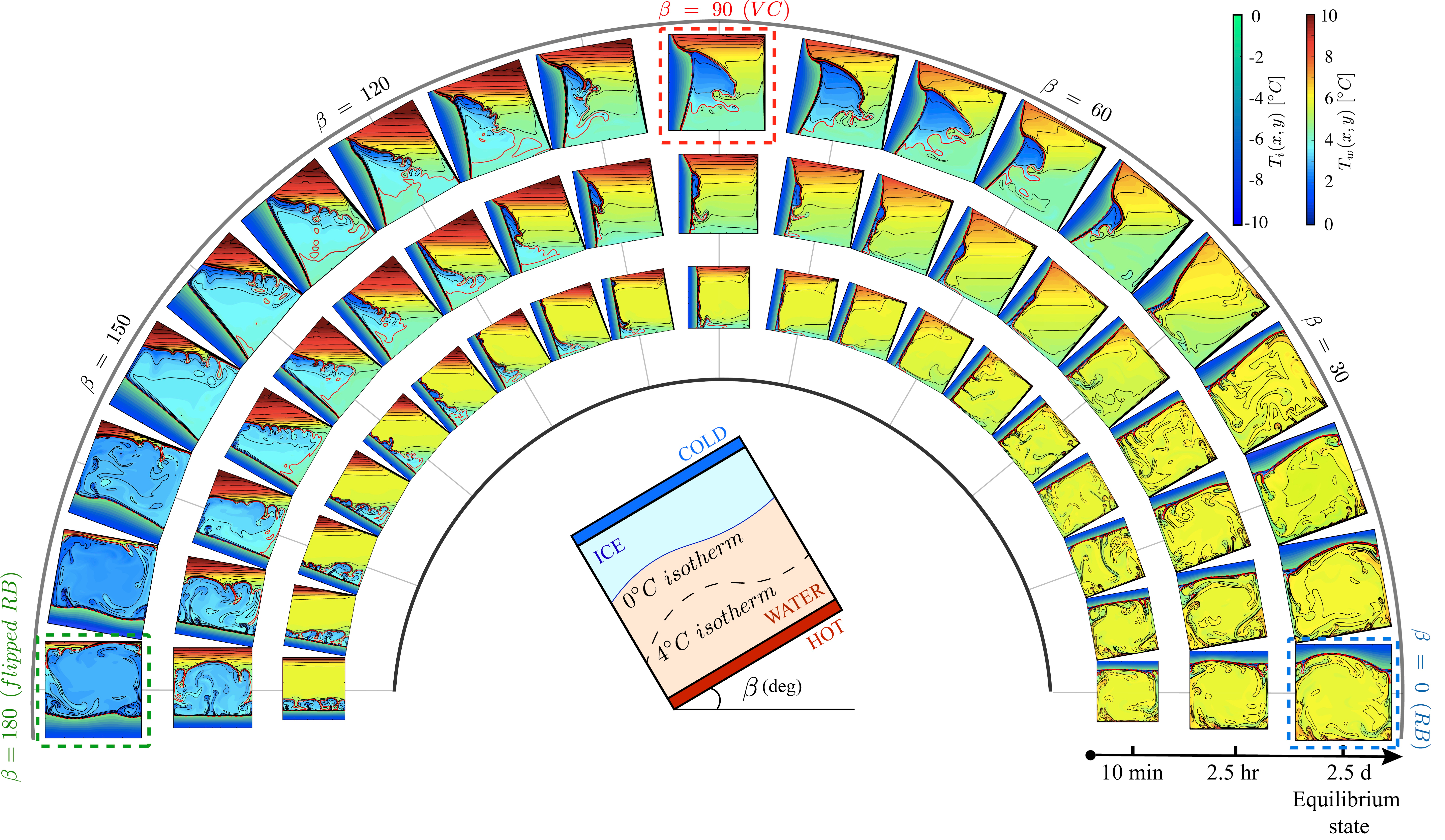}
			\caption{The systematic investigation of the influence of the inclination angle on the ice front morphology. The snapshots at the internal, middle, and outside circles are the temperature field 10min, 2.5hr, and 2.5d (when the system has reached the statistical equilibrium state) after the start of the simulations, respectively. The black thin lines are the isotherms. The red dashed line shows the ice front, and the red thick line shows the $T_c$-isotherm. The sketch in the middle shows the coordinate system (attached to the system) and how the inclination angle, $\beta$ (unit: deg), is defined.}
			\label{SM_systematic_angle} 
		\end{figure*}
		
		Fig.~\ref{SM_systematic_angle} shows the temperature field (overlapped with isotherms) with varying the system inclination angle, $\beta$ (unit: deg), from 0 (Rayleigh-B\'enard convection with heating from below and freshwater solidification from above) to $180$ (flipped Rayleigh-B\'enard convection with heating from above and freshwater solidification from below), and here we use the coordinate system attached to the cell. The snapshots at the internal, middle, and outside circles are the temperature field 10min, 2.5hr, and 2.5d (when the system has reached the statistical equilibrium state) after the start of the simulations, respectively. In the RB (small $\beta$) and the flipped-RB (large $\beta$) regimes, the ice front is heavily affected by the intensive convective flow, and thus the ice front represents complex forms of morphology different from those near the vertical convection cases. 
		
		%
		%
		
		\section{The boundary layer model and its input parameter}
		It has been mentioned that to locate the ice front profile, we need an input from the simulation results, namely the ice thickness at the starting point of the thermal boundary layer, $h_i(0)$. On top of this, the ice morphology can be predicted.
		
		Fig.~\ref{hinput} report the $h_i(0)$ as a function of the inclination angle, $\beta$. It can been observed that $h_i(0)$ is roughly constant and around the level of the averaged ice thickness $h_i$, which is shown in Fig.~3 of the main paper.
		\begin{figure*}[t]
			\centering
			\includegraphics[width=0.5\linewidth]{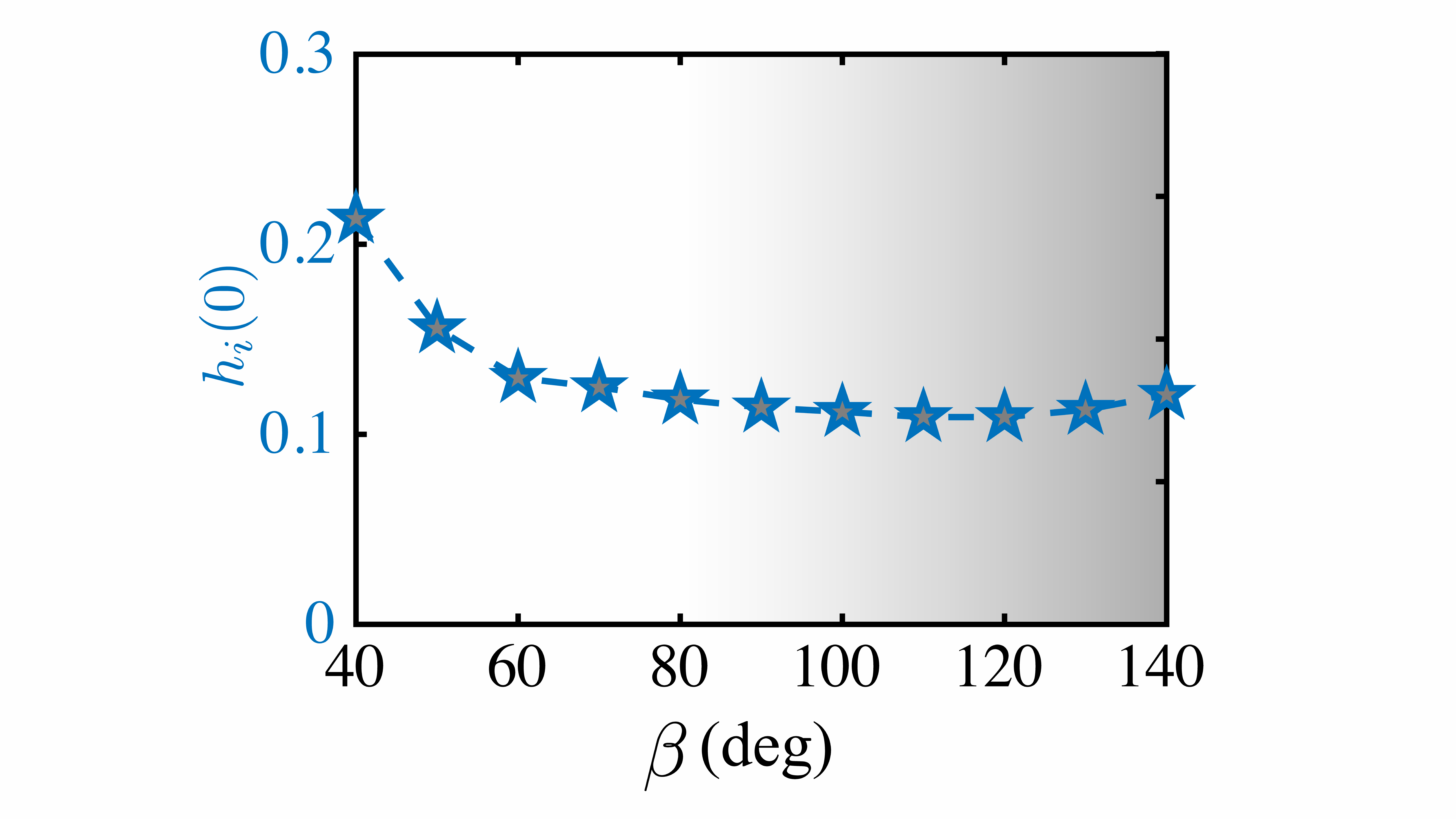}
			\caption{$h_i(0)$ as a function of the inclination angle, $\beta$. $h_i(0)$ is the ice thickness at $x=0$, which is the input parameter of the boundary layer model.}
			\label{hinput} 
		\end{figure*}
		
		{\color{black}
			\section{The buoyancy-intensity model}
			From the ice front morphology of the numerical simulation shown in Figure.4(c)-(m) of the main paper, we can observe that along x-axis, the ice thickness firstly increases and later decreases.  We indicate with $\Delta x_0$, the position where the ice thickness reaches the maximum (which is  here the a dimensionless position normalized by $H$).
			
			The prescribed system inclination results in different levels of thermal stratification, which induces modifications of the ice front morphology. When $\beta = 0$, 
			there is a stably-stratified layer (from $T_0$ to $T_c$) on top of the unstably-stratified layer (from $T_c$ to $T_b$). As the cell is tilted with a small $\beta$, the convection is strong enough to squeeze the stably-stratified layer to be closely attached to the ice front, so the ice morphology is influenced by the one-roll convective flow. As $\beta$ increases towards $90$, the inclination of the temperature gradient with respect to the gravity is strong enough to break down the stratification and thus the stably-stratified layer is set into motion in the form of a clockwise convective roll (an upward cold water current) which competes with the initially anticlockwise roll (downward warm water current). The shielding effect is prominent because the ice thickness reaches a local maximum, and the hot plumes impacting on the top part of the ice induces a local minimum of the ice thickness. The flow motion of the original stably-stratified layer (from $T_0$ to $T_c$) intensifies as $\beta$ increases but still when $\beta<90$. After $\beta=90$, the stratification configuration flips over.
			The temperature difference of the unstably-stratified layer is now $(T_c-T_0)\approx 4$K. The unstably stratified layer is directly in contact with the ice front.
			
			From the above described flow structure dependence on the inclination, we can deduce that there are two aspects of convection motion: 1) The intrinsic aspect, which means that even without inclination the convection exists, and this aspect of convective motion occurs in the initially unstably-stratified layer, i.e., from $T_c$ to $T_b$ when $\beta<90$ and from $T_0$ to $T_c$ when $\beta>90$. 2) The inclination-induced aspect, which means the initially stably-stratified layer can be set into convective motion when the inclination is introduced into the system. This aspect of convective motion occurs in the initially stably-stratified layer, i.e., from $T_0$ to $T_c$ when $\beta<90$ and from $T_c$ to $T_b$ when $\beta>90$. That's why in the definition of the buoyancy intensity, $I_1(\beta)$ and $I_2(\beta)$, there are two parts: the first part that includes $g_x$ results from the intrinsic aspect, and the second part that includes $g_y$ results from the inclination-induced aspect. 
			In order to display $I_1(\beta)$ and $I_2(\beta)$ in a more compact form, we introduce the Heaviside step function, $\mathcal{H}$, to turn on or off the intrinsic aspect. So the final definition of $I_1(\beta)$ and $I_2(\beta)$ is of this kind of form (see Eqn (\ref{buoyancyintensity}))
			
			\begin{equation}
				\begin{split}
					&I_1(\beta) =g_x (1- \tfrac{\rho(T_m)}{\rho_c}) - g_y \mathcal{H}[\beta-90^{\circ}] (1-\tfrac{\rho(T_0)}{\rho_c});\\
					&	I_2(\beta) =g_x (1-\tfrac{\rho(T_{m2})}{\rho_c}) + g_y \mathcal{H}[90^{\circ}-\beta] (1-\tfrac{\rho(T_b)}{\rho_c}).\\
				\end{split}
				\label{buoyancyintensity}
			\end{equation}
			
			The competition of the two convective rolls result in a local minimum heat transfer rate where the ice thickness reaches the maximum, and we use the buoyancy-intensity ratio to quantify this competition effect which give good agreement with the trend we observed in the numerical simulations (see Figure.4(b) of the main paper). This provides further evidence that the competition of two convective rolls accounts for the form of ice front morphology.\\

		}

\end{document}